\documentclass[12pt]{noHeadings}

\usepackage{hyperref}
\hypersetup{
    colorlinks=true,
    linkcolor=blue,
    filecolor=magenta,      
    urlcolor=cyan,
    pdftitle={Overleaf Example},
    pdfpagemode=FullScreen,
    }
\usepackage[capitalize,nameinlink,compress]{cleveref}
\usepackage{bm}
\usepackage{mathtools,amsmath}
\usepackage{caption,xcolor}
\usepackage{times}

\definecolor{darkgreen}{rgb}{0.0, 0.42, 0.24}

\newcommand{\blue}[1]{\textcolor{blue}{{#1}}}

\newcommand{\norm}[1]{\left\lVert#1\right\rVert}

\title[Distributed neural network control with dependability guarantees]{Distributed neural network control with dependability guarantees:\\ a compositional port-Hamiltonian  approach}

\author{%
 \Name{Luca Furieri} \Email{luca.furieri@epfl.ch}\\
 \addr \'Ecole Polytechnique Fédérale de Lausanne (EPFL), Institute of Mechanical Engineering, Switzerland
 \AND
 \Name{Clara Luc\'{i}a Galimberti} \Email{clara.galimberti@epfl.ch}\\
 \addr \'Ecole Polytechnique Fédérale de Lausanne (EPFL), Institute of Mechanical Engineering, Switzerland
 \AND
 \Name{Muhammad Zakwan} \Email{muhammad.zakwan@epfl.ch}\\
 \addr \'Ecole Polytechnique Fédérale de Lausanne (EPFL), Institute of Mechanical Engineering, Switzerland
 \AND
 \Name{Giancarlo {Ferrari-Trecate}} \Email{giancarlo.ferraritrecate@epfl.ch}\\
 \addr \'Ecole Polytechnique Fédérale de Lausanne (EPFL), Institute of Mechanical Engineering, Switzerland
}

\begin{document}
\maketitle
\begin{abstract}%
\footnote{The code for numerical examples is available at  \url{https://github.com/DecodEPFL/DeepDisCoPH}.}Large-scale cyber-physical systems require that control policies are \emph{distributed}, that is, that they only rely on local real-time measurements and communication with neighboring agents.  Optimal Distributed Control (ODC) problems are, however, highly intractable even in seemingly simple cases.  Recent work has thus proposed training Neural Network (NN) distributed controllers. A main challenge of NN controllers is that they are not \emph{dependable} during and after training, that is, the closed-loop  system may be unstable, and the training may fail due to vanishing and exploding gradients. In this paper, we address these issues for networks of nonlinear port-Hamiltonian (pH) systems,  whose modeling power ranges from energy systems to non-holonomic vehicles and chemical reactions. Specifically, we embrace the compositional properties of pH systems to characterize deep Hamiltonian control policies with \emph{built-in} closed-loop stability guarantees — irrespective of the interconnection topology and the chosen NN parameters. Furthermore, our setup enables leveraging recent results on well-behaved neural ODEs to prevent the phenomenon of vanishing gradients by design. Numerical experiments corroborate the dependability of the proposed architecture, while matching the performance of general neural network policies. 
\end{abstract}


\section{Introduction}
The design of distributed control policies — i.e., policies that can only observe local sensor measurements in real-time — can be extremely challenging even in simple cases.  Specifically, Witsenhausen's counter-example \citep{Witsenhausen} has shown that even when the system dynamics are \emph{linear}, the loss function to minimize is \emph{quadratic}, and the additive noise is \emph{Gaussian}, a \emph{nonlinear} distributed control policy can outperform the best linear one. A body of work culminating in \cite{rotkowitz2006characterization,lessard2011quadratic} has identified sufficient and necessary conditions — commonly known as Quadratic Invariance (QI) — under which optimal distributed control of linear systems is a convex (i.e., tractable) optimization program.   Nonetheless, most large-scale systems are nonlinear, and they violate the requirement of QI due to geographical distance or privacy concerns. These limitations motivate going beyond linear control and suggest parametrizing highly nonlinear distributed policies through Deep Neural Networks (DNNs). Specifically, the recent works \cite{tolstaya2020learning,khan2020graph,gama2021graph,yang2021communication} have focused on training Graph Neural Networks (GNNs) that parametrize static and dynamical distributed control policies. These methods achieve remarkable performance in applications such as vehicle flocking and formation flying. Furthermore, the special structure of GNNs allows for scalability to very large-scale systems.

The first challenge of using general GNNs as control policies is that it is often possible to guarantee closed-loop stability of the learned policy only under the assumptions that 1) the system dynamics are linear and open-loop stable, and 2) the DNN weights result in small-enough Lipschitz constants across the layers (e.g., \cite{gama2021graph}). As such conditions may not be fulfilled during exploration, the system may incur failures before an optimal policy can be learned \citep{brunke2021safe,cheng2019end}. Recent mitigation strategies include \cite{berkenkamp2018safe,richards2018lyapunov,koller2018learning}, where the authors propose to improve an initial known safe policy iteratively, while imposing the constraint that the initial \emph{region of attraction} does not shrink. An alternative approach  is to exploit integral quadratic constraints to enforce closed-loop stability of DNN controllers for linear systems \citep{pauli2021offset}.   The main obstacle of these methods is that explicitly constraining the DNN weights may be infeasible or become a bottleneck for closed-loop performance. Recent work proposes stable-by-design control policies based on mechanical energy conservation. However, the corresponding methods are specific to a class of robotic systems \citep{abdulkhader2021learning} and SE(3) dynamics \citep{duong2021hamiltonian}.

The second challenge of general $N$-layered GNN policies is that they must be applied recursively at every time step $t = 1,\ldots,T$. The corresponding optimal control problem is thus equivalent to a very deep learning problem involving $T\times N$ computational layers in total, which may exacerbate the phenomena of \emph{vanishing and exploding gradients}. Vanishing or exploding gradients may lead to numerical ill-posedness and impede the DNN training from learning a stabilizing controller.  Practical methods to mitigate exploding/vanishing gradients include clipping~\citep{GoodBengCour2016} and long short-term memory layers \citep{hochreiter1997long,pascanu2013difficulty}. A different approach is to design DNN architectures that avoid exploding/vanishing gradients \textit{by design}.  For instance, \cite{RN11052, RN11545, RN11544, choromanski2020ode} exploit unitary and orthogonal matrix trainable parameters. These approaches, however, require expensive computations during training  ~\citep{choromanski2020ode}, or use restricted classes of weight matrices~\citep{RN11052, RN11545, RN11544}, which may severely limit the expressivity of the models — especially so if combined with constraints for guaranteeing closed-loop stability. 

In this paper, we focus our attention on large-scale networks of nonlinear port-Hamiltonian (pH) systems.  Many large-scale engineering applications,  such as voltage and frequency control in islanded AC microgrids \citep{strehle2021unified}, swarms of robots \citep{angerer2017port}, and UAVs \citep{pH_Fixed,pH_Quad,pH_UAVS}, can be modeled as networks of coupled pH systems. Classical methods for stabilization of pH systems around desired equilibria include Passivity-Based Control (PBC)  and Control by Interconnection (CbI) \citep{ortega2008control,van2000l2}, and energy-based control via Casimir
functions and damping assignment, e.g.,  \cite{guo2006stabilization}. When it comes to minimizing a cost function, however, the problem becomes highly intractable. Therefore, only a few works address optimal control of nonlinear pH systems. For instance,   \cite{sprangers2014reinforcement} use reinforcement learning to include performance considerations into PBC, and \cite{kolsch2021optimal} incorporate an adaptive feedback component that steers the system towards an optimal solution for a modified cost. However, it has remained largely unexplored how to design distributed and deep neural network control policies that preserve closed-loop stability and good numerical properties during training. 

\paragraph{Contributions} 
We parametrize a novel class of distributed DNN control policies for large networks of pH systems that offers \emph{built-in guarantees} of closed-loop stability and non-vanishing gradients.  We achieve closed-loop stability — irrespective of the network topology and the choice of weights during exploration — by endowing Hamiltonian Deep Neural Networks (H-DNNs) from \cite{galimberti2021unified,galimberti2021hamiltonian} with input-output ports and exploiting compositional properties of pH systems. Simple sparsity assumptions on the trainable matrix weights of our controllers ensure that the policies are distributed, that is, that they comply with pre-defined information constraints typical of large-scale applications. 
We then show that the corresponding deep learning problem is not affected by the phenomenon of vanishing gradients when the pH dynamics are not dissipative.  We validate our results on a multi-robot collision avoidance task and we achieve the performance of general NN control policies.  To the best of our knowledge, no previous work offers guarantees of closed-loop and numerical stability  for distributed DNN control policies that hold irrespectively of the chosen weights and for a rich class of linear and nonlinear systems.

\paragraph{Notation}   
Let $\mathcal{G} = (\mathcal{V},\mathcal{E})$ be a graph with nodes $\mathcal{V} =\{1,\ldots,M\}$ and edges $\mathcal{E}$, and let $S\in\{0,1\}^{M \times M}$ be the associated adjacency matrix. The set of $k$-hop neighbors of node $i$ with respect to graph $\mathcal{G}$ is denoted as $\mathcal{N}_i^k$, that is, $\mathcal{N}_i^k = \{j \in \mathcal{V}|~~S^k(i,j) \neq 0\}$, where $S^k(i,j)$ denotes the entry $(i,j)$ of the matrix $S^k$. In the paper, we associate vectors $v_i \in\mathbb{R}^{n_i}$ with each node of graph $\mathcal{G}$.  The set of vectors associated with the $k$-hop neighbors of node $i$ with respect to graph $\mathcal{G}$ is denoted as $\mathbf{v}_i^k = \{v_j|~~j \in \mathcal{N}_i^k\}$. For a block-matrix $\mathbf{W} \in \mathbb{R}^{m \times n}$, where $m = \sum_{i=1}^M m_i$ and $n = \sum_{i=1}^N n_i$, we denote its block in position $(i,j)$ as $W_{ij} \in \mathbb{R}^{M_i \times n_i}$. For a binary matrix $R \in \{0,1\}^{M \times N}$, we write that $\mathbf{W} \in \operatorname{blkSparse}(R)$ if and only if {$R(i,j) = 0 \implies W_{ij} = 0$}. 


\section{Problem Statement}
We consider a network of $M \in \mathbb{N}$ nonlinear dynamically coupled systems, each endowed with a local feedback control policy. Letting $\mathcal{G}_d = (\mathcal{V}_d,\mathcal{E}_d)$ denote the graph associated with the dynamical couplings between systems, and $S_d\in \{0,1\}^{M \times M}$ with $S_d(i,i) = 1$ for all $i=1,\ldots,M$ be the adjacency matrix of $\mathcal{G}_d$, we have
\begin{align}
    \label{eq:system}
    \dot{x}_i(t) &= f_i(\mathbf{x}_i^1(t),u_i(t))\,, \\
    y_i(t) &= h_i(x_i(t))\nonumber\,, \qquad \forall i =1,\ldots,M\,,
\end{align}
where $x_i(t) \in \mathbb{R}^{n_i}$ is the local system state, $u_i(t) \in \mathbb{R}^{m_i}$ is the local control input, $y_i(t) \in \mathbb{R}^{p_i}$ is the local output, and $\mathbf{x}_i^1(t)$ denotes the set of states associated with $1$-hop neighbors of system $i$ according to $\mathcal{G}_d$. When the number $M$ of systems is very large, we say that system \eqref{eq:system} is \emph{large-scale}. The main challenge of control for large-scale systems is that the local control inputs $u_i(t)$ can only receive real-time information from a limited subset of neighboring systems according to a communication topology encoded by a graph $\mathcal{G}_c$ with adjacency matrix $S_c\in \{0,1\}^{M \times M}$ such that $S_c(i,i)=1$ for every $i=1,\ldots,M$. These policies are called \emph{distributed} due to the presence of information constraints. More formally, the goal of ODC is to compute optimal distributed feedback policies $(\chi_i(\cdot), \pi_i(\cdot))$ such that
\begin{align}
\dot{\xi}_i(t) &= \chi_i(\bm{\xi}_i^{R_\xi}(t), \mathbf{y}_i^{R_y}(t),t)\,,\label{eq:distributed_policy1}\\
u_i(t) &= \pi_i(\bm{\xi}_i^{R_\xi}(t), \mathbf{y}_i^{R_y}(t),t)\,, \quad \forall i=1,\ldots,M\,,\label{eq:distributed_policy2}
\end{align}
where $\bm{\xi}(t) = (\xi_1(t),\ldots,\xi_M(t))$, and $\mathbf{y}_i^{R_y}(t)$ and $\bm{\xi}_i^{R_\xi}(t)$ are defined in terms of the communication graph $\mathcal{G}_c$. In \eqref{eq:distributed_policy1}-\eqref{eq:distributed_policy2} the variable ${\xi}_i(t) \in \mathbb{R}^{q_i}$ is the internal dynamical state of the controller which acts as \emph{local memory}. In control-theoretic terms, \eqref{eq:distributed_policy1}-\eqref{eq:distributed_policy2} is a dynamical controller.  The value $R_y \in \mathbb{N}$ is the \emph{output measurement radius}, which indicates how far into the graph $\mathcal{G}_c$ local controllers are able to measure the outputs from other systems in real-time.  The value $R_\xi \in \mathbb{N}$, instead, is the \emph{communication radius}, which indicates how far into the graph $\mathcal{G}_c$ local controllers are able to exchange their local states $\xi_i(t)$ through communication in real-time. 

The policies \eqref{eq:distributed_policy1}-\eqref{eq:distributed_policy2} should be \emph{optimal} in the sense that they minimize a loss function
\begin{equation}
\label{eq:cost}
    \mathcal{L} = \frac{1}{T}\int_{0}^T l(\mathbf{x}(t),\mathbf{u}(t),t)dt\,,
\end{equation}
in a finite-horizon $T \in \mathbb{R}$, where $\mathbf{x}(t) = (x_1(t),\ldots,x_M(t))$ and $\mathbf{u}(t) = (u_1(t),\ldots,u_M(t))$.\footnote{The function $l(\cdot)$ may include an \emph{expected value} over a distribution on $\mathbf{x}(0)$. Our theoretical results are independent of this choice.}

We assume that $(\chi_i(\cdot),\pi_i(\cdot))$ are parametrized through tunable parameters $\theta_i(t) \in \mathbb{R}^{d_i}$ for every $t \in \mathbb{R}$ and $i=1,\ldots,M$. Therefore, the ODC amounts to finding optimal parameters $\bm{\theta}(t) = (\theta_1(t),\ldots,\theta_M(t))$ that solve the following optimization program:
\begin{alignat}{3}
&\min_{\bm{\theta}_{[0,T]}}~~~ &&\mathcal{L}(\bm{\theta}_{[0,T]})=\frac{1}{T} \int_{0}^T l(\mathbf{x}(t),\mathbf{u}(t),t)dt \label{eq:ODC_problem}\\
&\text{s.t.}&&\text{system dynamics \eqref{eq:system}}\,,\nonumber\\
&~&&\dot{\xi}_i(t) = \chi_i(\mathbf{y}_i^{R_y}(t),\bm{\xi}_i^{R_\xi}(t),\theta_i(t))\,,\label{eq:distr_contro_param1}\\
&~&&u_i(t) = \pi_i(\mathbf{y}_i^{R_y}(t),\bm{\xi}_i^{R_\xi}(t),\theta_i(t))\,, \quad \forall i=1,\ldots,M\,.\label{eq:distr_contro_param2}
\end{alignat}

When $l(\cdot)$ is quadratic in $\mathbf{x}(t)$ and $\mathbf{u}(t)$, the functions $f_i(\cdot)$ in \eqref{eq:system} are linear in their arguments,  and $R_y=R_\xi = M$, linear policies are optimal and \eqref{eq:ODC_problem} can be solved through convex or dynamic programming — this is the standard centralized LQ output-feedback problem. A similar result holds if $R_y$ and $R_\xi$ are strictly smaller than $M$, and the communication $\mathcal{G}_c$ is QI with respect to the dynamical couplings graph $\mathcal{G}_d$ of \eqref{eq:system} \citep{rotkowitz2006characterization}. Unfortunately, these assumptions do not hold for the vast majority of real-world large-scale systems. This fact motivates parametrizing the functions $\chi_i(\cdot),\pi_i(\cdot)$ as highly nonlinear deep neural networks, even when the dynamics $\eqref{eq:system}$ are linear \citep{gama2021graph,yang2021communication}.

However, the closed-loop interconnection of the system \eqref{eq:system} with the controller in the form \eqref{eq:distr_contro_param1}-\eqref{eq:distr_contro_param2} can lead to instability — that is, even if the cost is optimized for a finite-horizon $T$, the infinite-horizon cost is infinite, leading to system failure and safety concerns. Further, the corresponding learning problem may be ill-posed due to vanishing gradients when the horizon $T$ is large. In this work, we overcome these challenges for nonlinear system dynamics $\eqref{eq:system}$ that are in pH form.

\subsection{Networks of Port-Hamiltonian Systems}
A controlled network of nonlinear pH systems is a particular case of \eqref{eq:system} with
\begin{align}
    \dot{x}_i(t) &= (\Omega_i - R_i) \frac{\partial V_i(x_i(t))}{\partial x_i} + \sum_{j\in \mathcal{N}_i^1\setminus \{i\}}F_{ij}y_j(t)+G_i^\mathsf{T}u_i(t)\,,\label{eq:network_pH}\\
    y_i(t)&= G_{i}\frac{\partial V_i(x_i(t))}{\partial x_i}\,, \quad \forall i=1,\ldots,M\,,\nonumber
\end{align}
where ``$\setminus$'' denotes the set difference, $m_i = p_i$ for all $i=1,\ldots,M$, neighbors $j \in \mathcal{N}_i^1$ are defined in terms of the dynamical couplings graph $\mathcal{G}_d$, and it holds that $\Omega_i^\mathsf{T}+\Omega_i = 0$, $R_i\succeq 0$.  The functions $V_i(\cdot):\mathbb{R}^{n_i} \rightarrow \mathbb{R}$ are the local \emph{Hamiltonians} of the systems and  they describe their internal energy at any time $t \in \mathbb{R}$.
We make the following standard assumptions on the Hamiltonian functions:
\begin{enumerate}
    \item[\textbf{A1}] $V_i(\cdot)$ is continuously differentiable.
    \item[\textbf{A2}] $\lim_{\|x\|\rightarrow \infty}V_i(x) = \infty$, that is, $V_i(\cdot)$ is radially unbounded.
\end{enumerate}
The networked system \eqref{eq:network_pH} can be compactly written as
\begin{align}
    \dot{\mathbf{x}}(t) &= (\bm{\Omega}-\mathbf{R}) \frac{\partial V(\mathbf{x}(t))}{\partial \mathbf{x}}+\mathbf{F}\mathbf{y}(t)+\mathbf{G}^\mathsf{T}\mathbf{u}(t)\,, \label{network_1}\\
    \mathbf{y}(t)&= \mathbf{G} \frac{\partial V(\mathbf{x}(t))}{\partial \mathbf{x}}\,,\label{network_2}
\end{align}
where $\bm{\Omega} = \operatorname{blkdiag}(\Omega_i)$, $\mathbf{R} = \operatorname{blkdiag}(R_i)$, $\mathbf{F} \in \operatorname{blkSparse}(S_d)$ with $F_{ii}=0$ for every $i=1,\ldots,M$, $\mathbf{G} = \operatorname{blkdiag}(G_i)$ and $V(\mathbf{x}(t)) = \sum_{i=1}^MV_i(x_i(t))$. When the matrix $\mathbf{FG}$ is skew-symmetric, networks of pH systems in the form \eqref{eq:network_pH} possess a \emph{compositional property}, that is, their interconnection preserves stability properties of the local systems \citep{van2004port}. 
\begin{proposition}[Compositional principle for pH systems]
\label{prop:compositional}
Consider the network of pH systems in \eqref{network_1}-\eqref{network_2} with no external input $\mathbf{u}(t) = 0$, and assume that $\mathbf{FG}\in \operatorname{blkSparse}(S_d)$ is skew-symmetric. Then, we have that
\begin{equation}
\label{eq:convergence}
    \forall x_i(0) \in \mathbb{R}^{n_i}\,, \quad \lim_{t\rightarrow \infty}\left(\inf_{\overline{x}_i \in \mathcal{M}_i}\|x_i(t)-\overline{x}_i\|\right) = 0\,,\quad \forall i=1,\ldots,M\,,
\end{equation}
where $\mathcal{M}_i$ is the largest positively invariant set contained in  the set 
\begin{equation}
\label{eq:E_definition}
  E_i = \left\{\overline{x}_i|~~\frac{\partial V_i(\overline{x}_i)}{\partial x_i}\in \operatorname{Ker}(R_i)\right\}\,.
\end{equation}
\end{proposition}
We report the proof of Proposition~\ref{prop:compositional} in Appendix~\ref{app:CL_stability}. In plain words, Proposition~\ref{prop:compositional}  ensures that \emph{stability properties  of pH systems are not compromised by interconnecting arbitrarily many of them} through power-preserving links. Indeed, the states $x_i(t)$ will globally converge to the set $E_i$ — whose definition only depends on $V_i(\cdot)$ and $R_i$ — irrespective of dynamical couplings with neighboring pH systems $j \in \mathcal{N}_i^1$. Dynamical couplings can only affect the invariant sets $\mathcal{M}_i \subseteq E_i$ where the local state variables $x_i(t)$ converge to. We remark that while the result of Proposition~\ref{prop:compositional} holds with full generality, characterizing the invariant sets $\mathcal{M}_i$ requires explicit case-by-case analysis. For example, Proposition~\ref{prop:compositional} may also implies global asymptotic stability.
\paragraph{Example} If $R_i\succ 0$ for all $i=1,\ldots,M$, and the Hamiltonians $V_i(\cdot)$ are strongly convex, then $x_i(t)$ globally asymptotically converges to the unique stationary point $x_i^\star$ of $V_i(\cdot)$ for every $i=1,\ldots, M$, irrespective of the network topology and the dynamical couplings with other systems. This is because all $E_i$'s are singletons, and therefore, they are positively invariant.
 
 \vspace{0.5cm}
 
The compositional principle of Proposition~\ref{prop:compositional} inspires us to parametrize distributed DNN control policies that cannot compromise the stability properties of an arbitrarily large network of pH systems.
 

\section{De(e)pendable Distributed Control for Port-Hamiltonian Systems (DeepDisCoPH)}
We consider time-varying, dynamical control policies in the form
\begin{align}
    \dot{\bm{\xi}}(t) &= (\blue{\mathbf{J}}-\blue{\bm{R}_c})\frac{\partial \Phi(\bm{\xi}(t),\blue{\bm{\theta}(t))}}{\partial \bm{\xi}} +\blue{\mathbf{K}}\mathbf{y}(t)\,,
    \label{eq:controller1}\\
    \mathbf{u}(t)&= -\blue{\mathbf{K}^\mathsf{T}}\frac{\partial \Phi(\bm{\xi}(t),\blue{\bm{\theta}(t))}}{\partial \bm{\xi}}\,,\label{eq:controller2}
\end{align}
where explicit dependence on time is due to $\blue{\bm{\theta}(t)}$. In the equation above,  we assume that $\blue{\mathbf{J}^\mathsf{T}+\mathbf{J}}=0$ and $\blue{\mathbf{R}_c}\succeq 0$, where all trainable parameters are indicated in \blue{\textbf{blue color}}. The expressivity of control policies in the form \eqref{eq:controller1}-\eqref{eq:controller2} is linked to the scalar function $\Phi(\bm{\xi}(t),\blue{\bm{\theta}(t)})$, which parametrizes through $\blue{\bm{\theta}(t)}$ the \emph{internal energy} of the controller at any given time. We do not pose any restriction on the function $\Phi(\cdot)$ except from assumptions $\textbf{A1}$ and $\textbf{A2}$; for instance, $\Phi(\cdot)$ can be chosen as a simple quadratic function, as a Multi-Layer-Perceptron (MLP), or as a deep H-DNN \citep{galimberti2021hamiltonian}. The theoretical results of this paper hold irrespective of this choice.  

The class of policies \eqref{eq:controller1}-\eqref{eq:controller2} is chosen to structurally preserve the port-Hamiltonian form in the closed-loop. Our first observation is that, when the trainable parameters of the energy $\Phi(\cdot)$ do not depend on time, i.e., 
\begin{equation}
\label{eq:time_invariant_weights}
   \blue{\bm{\theta}(t)} = \blue{\bm{\theta}}\,,\quad \forall t \in \mathbb{R}\,,
\end{equation}
we can apply the compositional property of pH systems and achieve stability of the closed-loop \eqref{network_1}-\eqref{network_2}-\eqref{eq:controller1}-\eqref{eq:controller2} by design — i.e., irrespective of the specific values of $(\blue{\mathbf{J}}, \blue{\mathbf{R}_c}, \blue{\mathbf{K}}, \blue{\bm{\theta}})$. 
\begin{corollary}[Closed-loop stability by design]
\label{pr:CL_stability}
Consider the pH system \eqref{network_1}-\eqref{network_2} with skew-symmetric $\mathbf{FG}$  controlled through \eqref{eq:controller1}-\eqref{eq:controller2}. Assume that \eqref{eq:time_invariant_weights} holds. Then, for all $x_i(0)$ and all $i=1,\ldots, M$, the local state $x_i(t)$ converges to the largest invariant  set $\mathcal{M}_i$ contained in $E_i$, where $E_i$ is defined in \eqref{eq:E_definition}, and for all $\bm{\xi}(0)$, $\bm{\xi}(t)$ converges to the largest invariant set in $\left\{\overline{\bm{\xi}}|~\frac{\partial\Phi(\overline{\bm{\xi}},\blue{\bm{\theta}})}{\partial \bm{\xi}}\in \operatorname{Ker}(\blue{\mathbf{R}_c})\right\}$. 
\end{corollary}
The proof is equivalent to Proposition~\ref{prop:compositional} by noticing that the system \eqref{network_1}-\eqref{network_2} augmented with \eqref{eq:controller1}-\eqref{eq:controller2} is Hamiltonian with total energy $\sum_{i=1}^MV_i(x_i(t))+\Phi(\bm{\xi}(t),\blue{\bm{\theta}})$. We remark that the time-invariant assumption \eqref{eq:time_invariant_weights} is only needed to predict the infinite-time behavior of the closed-loop system. We do not require \eqref{eq:time_invariant_weights} during policy training because the performance metric $\mathcal{L}(\blue{\bm{\theta}_{[0,T]}})$ is defined in a finite-horizon. Hence, we suggest training a time-varying policy \eqref{eq:controller1}-\eqref{eq:controller2} for the time horizon of interest $[0,T]$ in order to preserve maximal expressivity, and then freeze the parameters $\bm{\theta}(t) = \blue{\bm{\theta}(T)}$ for all $t>T$ in order to preserve closed-loop stability according to Corollary~\ref{pr:CL_stability}.

\smallskip
The control policy \eqref{eq:controller1}-\eqref{eq:controller2} is \emph{not distributed}, in general. Indeed, the computation of a local control input $u_i(t)$ for some $i\in\{1,\ldots,M\}$ may involve output measurements $y_j(t)$ and controller internal states $\xi_k(t)$ of all the other systems. Our next result is to establish structural conditions on the trainable parameters and the energy function $\Phi(\cdot)$ to enforce a desired output measurement radius $R_y \in \mathbb{N}$ and communication radius $R_\xi \in \mathbb{N}$. The proof of Theorem~\ref{th:distributed} is reported in Appendix~\ref{app:distributed}.

\begin{theorem}[Distributed H-DNN controllers]
\label{th:distributed}
Consider the pH system \eqref{network_1}-\eqref{network_2} with $\mathbf{FG}$ skew-symmetric controlled through \eqref{eq:controller1}-\eqref{eq:controller2}. Assume that, for all $t \in \mathbb{R}$ and some $L_y,L_\xi \in \mathbb{N}$:
\begin{itemize}
    \item[\emph{\textbf{D1}}] the matrices $\blue{\mathbf{J}}$, $\blue{\mathbf{R}_c}$  and $\blue{\mathbf{K}}$ all belong to $\operatorname{blkSparse}((S_c)^{L_y})$,
    \item[\emph{\textbf{D2}}] the controller energy $\Phi(\cdot)$ is \emph{separable} as per
\begin{equation}
\label{eq:separable_energy}
 \Phi(\bm{\xi}(t),\blue{\bm{\theta}(t)}) = \sum_{i=1}^M \Phi_i(\bm{\xi}_i^{L_\xi}(t),\blue{\theta_i(t)})\,,
\end{equation}
that is, it is the sum of $M$ local energies, each one defined as a function of the internal state variables of the $L_\xi$-hop neighboring controllers according to the communication graph $\mathcal{G}_c$.
\end{itemize}
Then, the control policy \eqref{eq:controller1}-\eqref{eq:controller2} is distributed with output measurement radius $R_y$ and communication radius $R_\xi$ if
\begin{equation}
\label{eq:radius_condition}
    L_y \leq R_y,\text{ \emph{and} }~~ L_y+2L_\xi\leq R_\xi\,.
\end{equation}
\end{theorem}

\paragraph{Example} 
Consider the scenario where $R_y =R_\xi = 1$, corresponding to controllers that can only  measure outputs and exchange information with their one-hop neighbors according to the communication graph $\mathcal{G}_c$. We let $L_y=1$ and $L_\xi = 0$ to comply with the condition \eqref{eq:radius_condition}. Then, the assumption $\mathbf{D1}$ is satisfied by letting matrices $\blue{\mathbf{J}}$, $\blue{\mathbf{R}}_c$ and $\blue{\mathbf{K}}$ all lie in the subspace $\operatorname{blkSparse}(S_c)$. We satisfy assumption $\mathbf{D2}$ by letting the total controller energy $\Phi(\cdot)$ be  completely separable as per
\begin{equation*}
    \Phi(\bm{\xi}(t),\blue{\bm{\theta}(t)}) = \sum_{i=1}^M\Phi_i(\xi_i(t),\blue{\theta_i(t)})\,.
\end{equation*}
 For instance, letting $\blue{w_i}$ be a trainable row vector, one may choose the local energies as 
 \begin{equation}
     \label{eq:deeper_energy}
 \Phi_i(\xi_i(t),\blue{\theta_i}) = \blue{w_i} \sigma(\blue{W_{i,2}}\sigma(\blue{W_{i,1}}\xi_i(t)))\,, \quad \forall i=1,\ldots, M\,,
 \end{equation}
 which corresponds to the output of a two-layer NN with activation function $\sigma(\cdot)$.


\section{Training DeepDisCoPH Controllers Beyond Vanishing Gradients}
Having established a class of distributed control policies that preserves closed-loop stability by design, we now seek a low-cost stationary point for \eqref{eq:ODC_problem} by training a Neural ODE. 
For details on its implementation, we refer the interested reader to \cite{NeuralODEs}. 
In summary,  neural ODEs are neural network models that generalize standard layer to layer propagation to continuous-time dynamics. This is achieved by interpreting the function $\mathbf{f}(\cdot)$ of an ODE $\dot{\mathbf{x}}(t) = \mathbf{f}(\mathbf{x}(t),\blue{\bm{\theta}(t)})$ as a NN  endowed with an integration method. We remark that discretization during training can introduce suboptimality, but does not compromise the closed-loop stability result of Corollary~\ref{pr:CL_stability}; indeed, stability of the continuous-time systems holds irrespectively of the chosen weights.

Next, we  derive the neural ODE model associated with the trainable closed-loop system \eqref{network_1}-\eqref{network_2} controlled through \eqref{eq:controller1}-\eqref{eq:controller2}. Then, we establish a property of non-vanishing gradients during training for non-dissipative systems  — irrespective of the neural ODE depth. 

\paragraph{ODC as an H-DNN}
The neural ODE associated with our class of pH systems and controllers is expressed as

\begin{align}
\dot{\bm{\zeta}}(t)=\begin{bmatrix}\dot{\mathbf{x}}(t)\\\dot{\bm{\xi}}(t)\end{bmatrix} &= \begin{bmatrix}\bm{\Omega}+\mathbf{FG}-\mathbf{R}&-\mathbf{G}^\mathsf{T}\blue{\mathbf{K}}^\mathsf{T}\\
\blue{\mathbf{K}}\mathbf{G}&(\blue{\mathbf{J}}-\blue{\mathbf{R}_c})\end{bmatrix}\begin{bmatrix}
\frac{\partial V(\mathbf{x}(t))}{\partial \mathbf{x}}\\ \frac{\partial \Phi(\bm{\xi}(t),\blue{\bm{\theta}(t)})}{\partial \bm{\xi}}
\end{bmatrix}\label{eq:closed_Loop_footnote}\\
&=(\blue{\bm{\Psi}}-\blue{\mathbf{S}})\frac{\partial P(\bm{\zeta}(t),\blue{\bm{\theta}(t)})}{\partial \bm{\zeta}}\,,\label{eq:closed_Loop}
\end{align}
where $\blue{\bm{\Psi}}$ is  skew-symmetric and $\blue{\mathbf{S}}\succeq 0$ for every $t\in \mathbb{R}$.\footnote{Note that both $\blue{\bm{\Psi}}$ and $\blue{\mathbf{S}}$ contain both trainable (in \blue{\textbf{blue}}) and non-trainable (in \textbf{black}) parameters as per \eqref{eq:closed_Loop_footnote}. We choose nevertheless to represent them in \textbf{\textcolor{blue}{blue}}.} 
The Hamiltonian function of the closed-loop system is expressed as $P(\mathbf{x}(t),\bm{\xi}(t),\blue{\bm{\theta}(t)})= V(\mathbf{x}(t))+\Phi(\bm{\xi}(t),\blue{\bm{\theta}(t)})$. 
The corresponding neural ODE has the form of a time-varying Hamiltonian system — as recently proposed in \cite{galimberti2021hamiltonian}. 
For instance, upon implementing a Forward Euler (FE) discretization scheme for \eqref{eq:closed_Loop}, the corresponding neural network architecture is akin to  H$_1$-DNN from \cite{galimberti2021hamiltonian}. 
Here, the main difference with respect to \cite{galimberti2021hamiltonian} is that only the controller parameters highlighted in blue in \eqref{eq:closed_Loop_footnote} can be trained through back-propagation, and that the performance metric considers the entire trajectory $\bm{\zeta}_{[0,T]}$ rather than  $\bm{\zeta}(T)$ only. 
Further, the energy $\Phi(\cdot)$ is not restricted to a fixed model, but can be chosen arbitrarily, e.g. \eqref{eq:deeper_energy}. 
Irrespective of this choice, we show that the Backward Sensitivity Matrices (BSMs) cannot vanish.

\allowdisplaybreaks
\begin{theorem}[Non-vanishing sensitivities]
\label{th:vanishing}
Consider the ODC problem \eqref{eq:ODC_problem}. Assume that the networked system and the local control policies are in pH form, i.e. \eqref{network_1}-\eqref{network_2} with $\mathbf{FG}$ skew-symmetric and \eqref{eq:controller1}-\eqref{eq:controller2} hold.  Further assume that $\bm{R} = \blue{\bm{R}_c} = 0$. Then, 
\begin{equation*}
    \norm{\frac{\partial {\bm{\zeta}}(\tilde{t})}{\partial {\bm{\zeta}}(\tilde{t}-t)}} \geq 1\,, \quad \forall t,\tilde{t}, T \in \mathbb{R}, \text{\emph{ such that} } t\leq \tilde{t}\leq T\,,
\end{equation*}
where $\bm{\zeta}(t) = (\mathbf{x}(t),\bm{\xi}(t))$, and $\norm{\cdot}$ is any sub-multiplicative norm with $\norm{X^\mathsf{T}} = \norm{X}$.
\end{theorem}
The proof of Theorem~\ref{th:vanishing} is provided in Appendix~\ref{app:vanishing}. As it was shown in \cite{galimberti2021hamiltonian}, the fact that the BSM $\frac{\partial {\bm{\zeta}}(\tilde{t})}{\partial {\bm{\zeta}}(\tilde{t}-t)}$ cannot vanish as $(\tilde{t}-t)$ increases plays a key role in guaranteeing non-vanishing gradients for arbitrarily deep H-DNN implementations of the ODC \eqref{eq:ODC_problem}. Indeed, letting $\blue{\theta_{i,j}}$ denote the scalar parameter $i$ of layer $j$ of the $N$-layered neural ODE implementation, the backward propagation algorithm performs the following computations
\begin{align}
\frac{\partial \mathcal{L}}{\partial \blue{\theta_{i,j}}} = \frac{\partial \bm{\zeta}_{j+1}}{\partial \blue{\theta_{i,j}}} \frac{\partial \mathcal{L}}{\partial \bm{\zeta}_{j+1}}  =  \frac{\partial \bm{\zeta}_{j+1}}{\partial \blue{\theta_{i,j}}}\sum_{k=j+1}^N \left[ 
\left(  \frac{\partial \bm{\zeta}_{k}}{\partial \bm{\zeta}_{j+1}} \right)
\frac{\partial \mathcal{L}}{\partial \bm{\zeta}_{k}}\right]\,.\label{eq:parameterGradient}
\end{align}
Vanishing (resp., exploding) gradients are linked to the term $\frac{\partial \bm{\zeta}_{k}}{\partial \bm{\zeta}_{j+1}}$ vanishing to zero (resp., diverging to infinity) as $k \rightarrow N$ and $N\rightarrow \infty$. Since $\frac{\partial \bm{\zeta}_{k}}{\partial \bm{\zeta}_{j+1}}$ is the discrete-time counterpart of $\frac{\partial {\bm{\zeta}}(\tilde{t})}{\partial {\bm{\zeta}}(\tilde{t}-t)}$, our result of Theorem~\ref{th:vanishing} ensures that the proposed architecture prevents vanishing gradients in continuous-time, and may only appear due to $1)$ an excessively large discretization step  in the neural ODE implementation, or $2)$ the resistive part of pH systems.


\section{Numerical experiments}
\allowdisplaybreaks
In this section, we validate our results on  dependability and performance of DeepDisCoPH control policies on an ODC task. We consider a fleet of  $M=12$ mobile robots that need to achieve a formation on the $xy$-plane within a given time $T \in \mathbb{R}$. 
The robots can  only rely on neighbors' information, must avoid collisions between each other, and their trajectories should minimize a given loss function. Each robot $i=1,\ldots,M$ is modeled through point-mass dynamics with mass $m_i>0$, and the control input $\mathbf{u}(t) = (u_1(t),\ldots,u_M(t))$ is the force that local motors can apply to the corresponding robot. 
The state of robot $i$ is given by  $x_i(t) = (q_{i,x}(t), q_{i,y}(t), p_{i,x}(t), p_{i,y}(t))$, where $(q_{i,x}(t),q_{i,y}(t))$ and $(p_{i,x}(t), p_{i,y}(t))$ are the position and momentum  of the robot at time $t \in \mathbb{R}$, respectively.  The robots possess a guidance law that steers them to their target positions $\overline{x}_i$ — albeit, without avoiding collisions nor maximizing a performance metric.  
The dynamics are pH with 
\begin{equation*}
    V(\mathbf{x}(t)) = \sum_{i=1}^MV_i(x_i(t)), \quad V_i(x_i(t)) = (x_i(t)-\bar{x}_i)^\mathsf{T}U_i(x_i(t)-\bar{x}_i)\,,
\end{equation*}
where $U_i = \operatorname{blkdiag}(\frac{1}{2m_i}I_2,\frac{k_i}{2}I_2)$ and $k_i\geq 0$ represents the virtual spring of the initial guidance law. 
We defer the detailed expression of the pH system to Appendix~\ref{app:implementation}. As the robots are not dynamically coupled, the graph $\mathcal{G}_d$ only has self-loops and thus $S_d = I_{12}$.

\paragraph{Information constraints} 
We assume that robot $i$ can only receive information in real-time from robot $i-1$ and robot $i+1$\footnote{Technically, $\mathbf{mod}(i-1,M)$ and $\mathbf{mod}(i+1,M)$, where $\mathbf{mod}(\cdot)$ indicates the modulo operation.} according to a circular communication graph $\mathcal{G}_c$ with adjacency matrix $S_c$ reported in \eqref{eq:communication_matrix} of Appendix~\ref{app:implementation}. 
Hence, we require output measurement and communication radii of $R_y=R_\xi = 1$. 
According to Theorem~\ref{th:distributed}, we set $L_y = 1$ and $L_\xi = 0$ to satisfy \eqref{eq:radius_condition}. Accordingly, we impose $\blue{\mathbf{J}},\blue{\mathbf{R}_c}, \blue{\mathbf{K}} \in \operatorname{blkSparse}(S_c)$ and select a fully separable $\Phi(\bm{\xi}(t)) = \sum_{i=1}^M\Phi_i(\xi_i(t),\blue{\theta_i(t)})$. 
For this experiment, 
we choose the controller energy as in \cite{galimberti2021hamiltonian} given by 
\begin{equation}
\label{eq:energy_example}
   \Phi(\bm{\xi}(t),\blue{\bm{\theta}}) = \sum_{i=1}^M \Phi_i(\xi_i(t),\blue{\theta_i(t)}) = \sum_{i=1}^M\operatorname{log}\left(\operatorname{cosh}\left(\blue{W_i(t)}\xi_i(t)+\blue{b_i(t)}\right)\right)^\mathsf{T}\mathbf{1}\,.
\end{equation}

\paragraph{Cost function and collision avoidance}
The loss function is composed of three main addends. The first addend, denoted as $\mathcal{L}_x$, is the standard control cost that penalizes the overall distance of robots from their targets and the control effort over a time horizon $T \in \mathbb{R}$, and is given by
\begin{equation}
\label{eq:Lx}
	\mathcal{L}_{x}
	= 
	\int_0^T \left[(\mathbf{x}(t)-\bar{\mathbf{x}})^\top \mathbf{Q}(t)\, (\mathbf{x}(t)-\bar{\mathbf{x}})^\top + \mathbf{u}^\top(t) \mathbf{R}(t) \, \mathbf{u}(t)\right]  \, dt \,.
\end{equation}
The second addend, denoted as $\mathcal{L}_{ca}$, strongly penalizes the event that any two robots $i$ and $j$ find themselves at a distance $d_{ij}(t) < D$, where $D \in \mathbb{R}$ is a pre-defined safety distance. Accordingly, 
\begin{equation}
\label{eq:Lca}
\mathcal{L}_{ca} 
	= 
	\int_0^T \sum_{i=1}^{M-1} \sum_{j=i+1}^{M}L_{ca}^{ij}(t) \, dt\,,
\end{equation}
 where $L_{ca}^{ij}(t)
	=  (d_{ij}(t)+\epsilon)^{-2}$ if $d_{ij}(t) \leq D$, and $L_{ca}^{ij}(t) = 0$ otherwise,  
and $\epsilon>0$ is a small constant such that $L_{ca}^{i,j}(t)<\infty$ at all times. 
The third addend of the loss function is a regularization term $\mathcal{R}_{w} = \int_{0}^{T} \sum_{i=1}^M\norm{\blue{W_i(t+\Delta )}-\blue{W_i(t)}}^2 dt$ that encourages smooth weight variations in time, and $\Delta>0$ is a small value that acts as a discretization step-size. Overall, the total loss function is expressed as $\mathcal{L} = \mathcal{L}_x + \alpha_{ca}\mathcal{L}_{ca} + \alpha_w \mathcal{R}_w$, where $(\alpha_{ca},\alpha_w)>0$ are hyperparameters.

\begin{figure}
    \begin{minipage}{0.5\linewidth}
        \includegraphics[width=\linewidth]{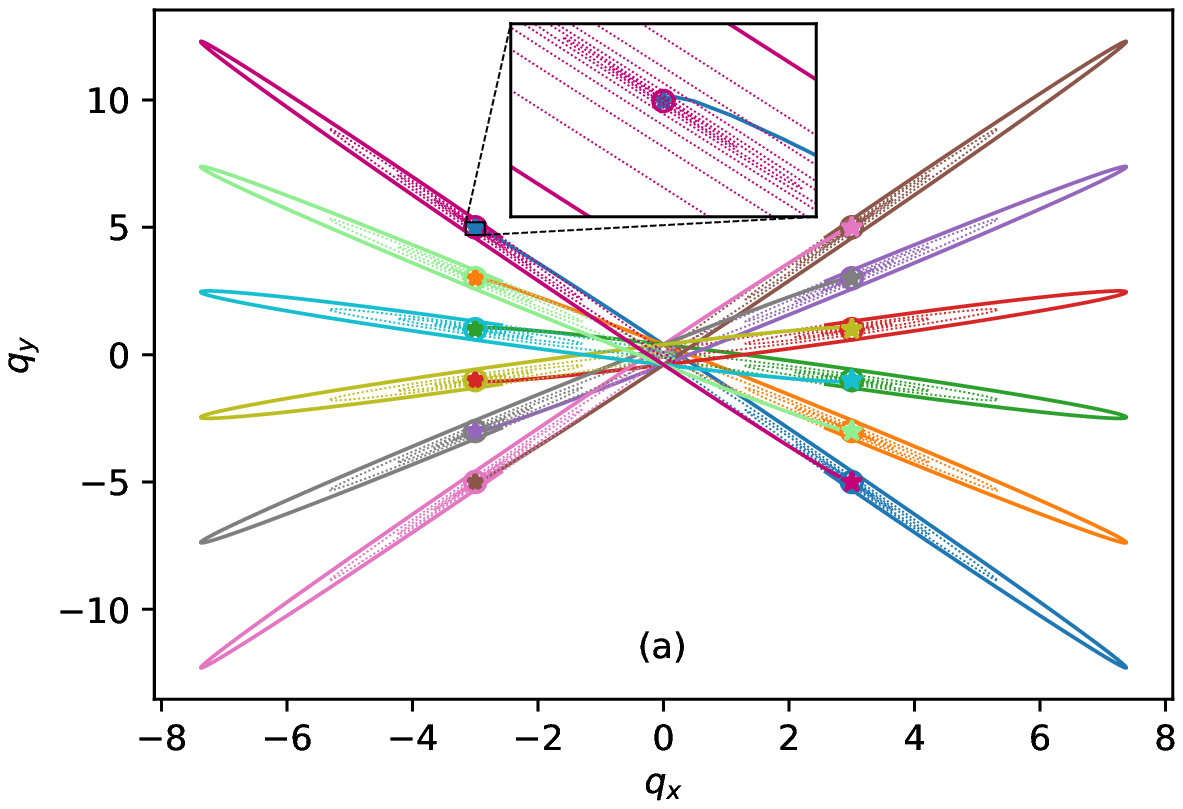}
    \end{minipage}%
    \begin{minipage}{0.5\linewidth}
        \includegraphics[width=\linewidth]{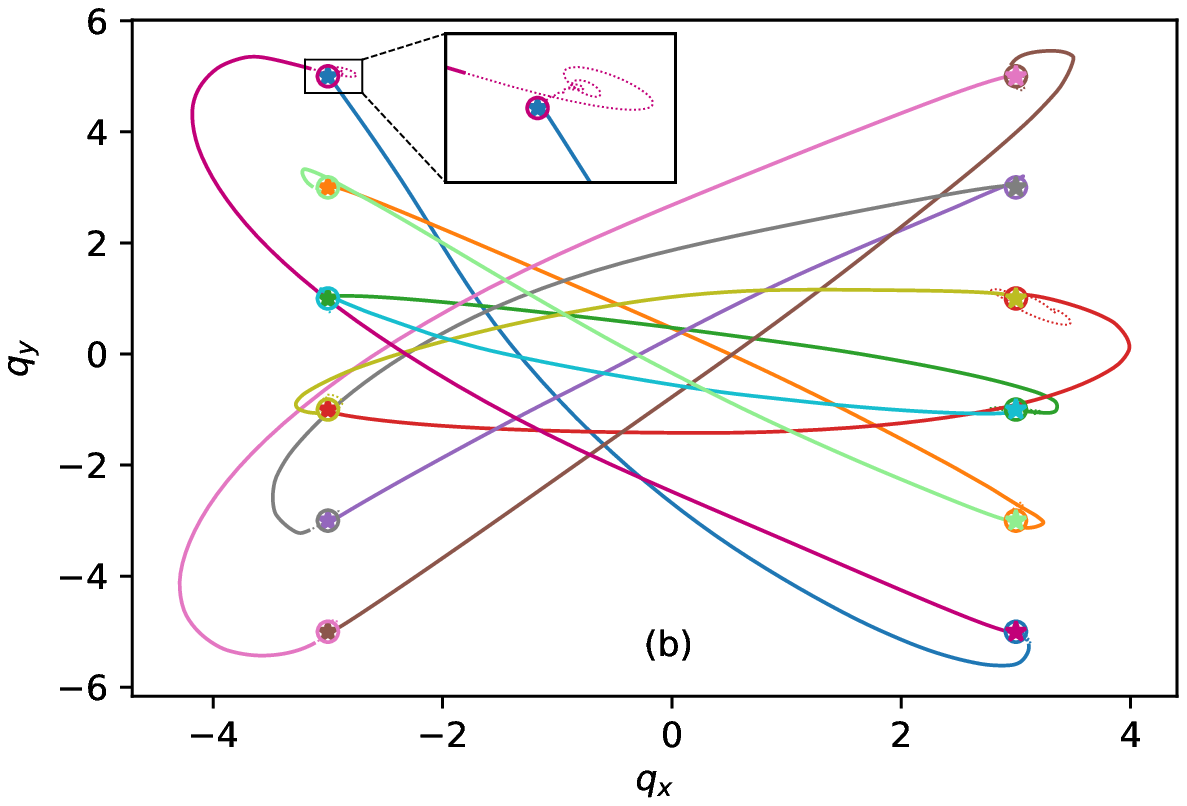}
    \end{minipage}
    \caption{Trajectories of the robots in the $xy$-plane, (a) before training, and (b) after training DeepDisCoPH controllers. Dotted lines indicate the trajectories for $t>T$, where $T$ is the training horizon. Target positions indicated with $\circ$. Initial positions indicated with $\star$.}
 \label{fig:OL_CL}
\end{figure}

Figure~\ref{fig:OL_CL}(a) shows robot trajectories before training; target positions are not reached within $T = 5$ seconds and multiple collisions occur around $(0,0)$. Figure~\ref{fig:OL_CL}(b) shows trajectories after training DeepDisCoPH controllers \eqref{eq:controller1}-\eqref{eq:controller2} with energy parametrized as per \eqref{eq:energy_example}.  The target positions $\circ$ are reached within the time horizon of $T= 5$ seconds and collisions are avoided while also optimizing the control cost $\mathcal{L}_x$. Videos associated with Figure~\ref{fig:OL_CL} are available \href{https://github.com/DecodEPFL/DeepDisCoPH}{here}. We note that distributed feedback policies favor robustness. Using the trained weights for perturbed initial states sampled  from a radius $r=0.5$ around $\mathbf{x}(0)$, collisions are  avoided in $95\%$ of the runs. 

\paragraph{Comparison with distributed MLPs \citep{gama2021graph}} 
We compare the closed-loop stability properties of DeepDisCoPH controllers with a trained distributed MLP controller as per  \cite{gama2021graph}.\footnote{Without the structural assumptions needed for a scalable GNN implementation.} We refer the reader to Appendix~\ref{app:mlp} for full details and results. Upon simulating the closed-loop system for a long horizon of $10T$,  a trained MLP may destabilize the system for $t>T$. Instead, a DeepDisCoPH controller preserves stability as predicted by Corollary~\ref{pr:CL_stability}. For a fair comparison with distributed MLP controllers, we have also trained a \textit{time-invariant} DeepDisCoPH controller with a comparable number of training parameters, and observed the same phenomenon after training. 

\paragraph{Non-vanishing gradients}
We computed the norms of the terms $\frac{\partial \bm{\zeta}_j}{\partial \bm{\zeta}_i}$ for all $i,j$ satisfying $0\leq i < j \leq N$ at every iteration, and verified that gradients never tend to vanish, despite a high number of neural ODE layers and dissipation in the dynamics. We refer the reader to Appendix~\ref{app:gradients} for details.

\paragraph{Dependability during training} 
We also verify that DeepDisCoPH controllers preserve closed-loop stability despite prematurely stopping of  the training, as predicted by Corollary~\ref{pr:CL_stability}. Our results are reported in Appendix~\ref{app:early_stop}, where we plot the closed-loop trajectories of the robots for a long time horizon, upon applying a DeepDisCoPH controller which has been trained for 5\%, 25\%, 50\% and 75\% of the epochs that are necessary to achieve the performance of Figure~\ref{fig:OL_CL}.


\section{Conclusions}
As we move towards highly nonlinear deep distributed control for safety-critical cyberphysical systems, dependability during exploration becomes a primary concern. We have proposed deep, and dependable, distributed control policies that preserve closed-loop stability and non-vanishing gradients for arbitrarily large networks of nonlinear pH systems (DeepDisCoPH).   Near-optimal performance can be achieved thanks to parametrizing deep nonlinear Hamiltonian energy functions for the controllers. 


\acks{Research supported by the Swiss NSF under the NCCR Automation (grant agreement 51NF40\textunderscore 180545).}
\bibliography{references}

\newpage


\appendix

\section{Proof of Proposition~\ref{prop:compositional}}
\label{app:CL_stability}
When $\mathbf{u}(t)=0$, the networked system can be rewritten as
\begin{equation}
\label{eq:network_compact}
\dot{\mathbf{x}}(t) = (\bm{\Omega}+\mathbf{FG}-\mathbf{R})\frac{\partial V(\mathbf{x}(t))}{\partial \mathbf{x}}|_{\mathbf{x}=\mathbf{x}(t)}\,.
\end{equation}
 If $\mathbf{FG}$ is skew-symmetric, we have
\begin{align}
    \dot{V}(\mathbf{x}(t)) &= \frac{\partial V(\mathbf{x}(t))}{\partial \mathbf{x}} \dot{\mathbf{x}}(t) = -\frac{\partial V(\mathbf{x}(t))}{\partial \mathbf{x}}^\mathsf{T} \mathbf{R}\frac{\partial V(\mathbf{x}(t))}{\partial \mathbf{x}}\nonumber\\
    &= -\sum_{i=1}^M\frac{\partial V_i(x_i(t))}{\partial x_i}^\mathsf{T}R_i\frac{\partial V_i(x_i(t))}{\partial x_i}\leq 0\,.\label{dissipation}
\end{align}
Since $V_i(\cdot)$ is radially unbounded for every $i=1,\ldots,M$, then $V(\cdot)$ is radially unbounded. Hence, all level sets of $V(\cdot)$ are bounded, and since  $V(\mathbf{x}(0))< \infty$ and $\dot{V}(\mathbf{x}(t))\leq 0 $ for every $t$ all system trajectories are bounded. By Krasovski-LaSalle's Invariance Principle \citep{khalil2002nonlinear} we conclude that, for any initial state  $\mathbf{x}(0)$, the state $\mathbf{x}(t)$ converges to the largest invariant set contained in 
\begin{align*}
E &= \left\{\overline{\mathbf{x}}|~\frac{\partial V(\overline{\mathbf{x}})}{\partial \mathbf{x}}^\mathsf{T}\mathbf{R}\frac{\partial V(\overline{\mathbf{x}})}{\partial \mathbf{x}}= 0\right\}\\
&= \left\{\overline{\mathbf{x}}|~\frac{\partial V_i(\overline{x}_i)}{\partial x_i}^\mathsf{T}R_i \frac{\partial V_i(\overline{x}_i)}{\partial x_i} = 0,~~\forall i=1,\ldots,M\right\}\\
&= E_1 \times E_2 \times \ldots \times E_M\,, 
\end{align*}
where, for every $i=1,\ldots,M$, it holds that $E_i = \left\{\overline{x}_i|~\frac{\partial V_i(\overline{x}_i)}{\partial x_i} \in \operatorname{Ker}\left(R_i\right)\right\}$.


\section{Proof of Theorem~\ref{th:distributed}}
\label{app:distributed}
Let $\blue{J_{ij}}$, $\blue{R_{c,ij}}$  and $\blue{K_{ij}}$ denote the blocks located at position $(i,j)$ of $\blue{\mathbf{J}}$, $\blue{\mathbf{R}_c}$ and $\blue{\mathbf{K}}$, respectively. For any $i \in \{1,\ldots,M\}$, the local controller equation can be written as
\begin{align*}
    \dot{\xi}_i(t) &= \sum_{j \in \mathcal{N}_i^{L_y}}\left[(\blue{J_{ij}}-\blue{R_{c,ij}})\frac{\partial \left(\sum_{l \in \mathcal{N}_j^{L_\xi}}\Phi_l\left(\bm{\xi}_l^{L_\xi}(t),\blue{\theta_l(t)}\right)\right)}{\partial \xi_j}+\blue{K_{ij}}y_j(t)\right]\,,\\
    u_i(t) &= -\sum_{j \in \mathcal{N}_i^{L_y}}\left[(\blue{K_{ji}})^\mathsf{T}\frac{\partial \left(\sum_{l \in \mathcal{N}_j^{L_\xi}}\Phi_l\left(\bm{\xi}_l^{L_\xi}(t),\blue{\theta_l(t)}\right)\right)}{\partial \xi_j}\right]\,.
\end{align*}
By inspection, we conclude that $\dot{\xi}_i(t)$ is computed as a function of the internal controller variables $\xi_l(t)$ of at most its $(L_y+2L_\xi)$-hops neighbors and the output measurements of at most its $L_y$-hops neighbors. Further, the local control input $u_i(t)$ is computed as a function of the internal controller variables $\xi_l(t)$ of at most its $(L_y+2L_\xi)$-hops neighbors. The condition \eqref{eq:radius_condition} follows. 


\section{Proof of Theorem~\ref{th:vanishing}}
\label{app:vanishing}
 Let $Z\triangleq \frac{\partial {\bm{\zeta}}(\tilde{t})}{\partial {\bm{\zeta}}(\tilde{t}-t)}$ for brevity and $g(\bm{\zeta}(t),\blue{\bm{\theta}(t)}) = \bm{\Psi}\frac{\partial P(\bm{\zeta}(t),\blue{\bm{\theta}(t)})}{\partial \bm{\zeta}}$. By adapting Lemma~1 of \cite{galimberti2021hamiltonian}, we have that
 \begin{equation}
     \dot{Z} = \frac{\partial^\mathsf{T} g(\bm{\zeta}(\tilde{t}-t),\blue{\bm{\theta}(\tilde{t}-t)})}{\partial \bm{\zeta}} Z\,.\label{eq:BSM}
 \end{equation}
 By \eqref{eq:closed_Loop}, \eqref{eq:BSM} and $\bm{\Psi} = -\bm{\Psi}^\mathsf{T}$ we obtain
 \begin{align*}
     &\frac{d}{dt}\left(Z^\top\bm{\Psi}Z\right) = \dot{Z}^\top\bm{\Psi}Z +Z^\top\bm{\Psi}\dot{Z}\\
     &=Z^\mathsf{T} \frac{\partial g(\bm{\zeta}(\tilde{t}-t),\blue{\bm{\theta}(\tilde{t}-t)})}{\partial \bm{\zeta}} \bm{\Psi}Z+Z^\mathsf{T} \bm{\Psi} \frac{\partial^\mathsf{T} g(\bm{\zeta}(\tilde{t}-t),\blue{\bm{\theta}(\tilde{t}-t)})}{\partial \bm{\zeta}} Z\\
     &=Z^\mathsf{T}\left(\bm{\Psi}\frac{\partial^2 P(\bm{\zeta}(\tilde{t}-t),\blue{\bm{\theta}(\tilde{t}-t)})}{\partial \bm{\zeta}^2}\bm{\Psi}+\bm{\Psi}\frac{\partial^2 P(\bm{\zeta}(\tilde{t}-t),\blue{\bm{\theta}(\tilde{t}-t)})}{\partial \bm{\zeta}^2}\bm{\Psi}^\mathsf{T}\right)Z = 0\,.
 \end{align*}
 Since $\frac{\partial {\bm{\zeta}}(\tilde{t})}{\partial {\bm{\zeta}}(\tilde{t})} = I$ and the quantity $Z^\mathsf{T}\bm{\Psi}Z$ is constant over time, we deduce that
 \begin{equation*}
     \norm{\bm{\Psi}} = \norm{Z^\mathsf{T}\bm{\Psi}Z}\leq \norm{Z}^2\norm{\bm{\Psi}}\,,
 \end{equation*}
and hence $\norm{Z}\geq 1$ for every $t$ and $\tilde{t}$ such that $t\leq \tilde{t} \leq T$ for any $T \in \mathbb{R}$.


\section{Implementation}
\label{app:implementation}

We consider a fleet of $M=12$ mobile robots that need to achieve a pre-specified formation described by $(\bar{q}_{i,x}, \bar{q}_{i,y})$ for each agent $i$ and with zero velocity.
Each vehicle $i$ endowed with a guidance law is described by
\begin{equation}
\begin{bmatrix}
	\dot{p}_{i,x}\\
	\dot{p}_{i,y}\\
	\dot{q}_{i,x}\\
	\dot{q}_{i,y}
\end{bmatrix}
=
\left(
\begin{bmatrix}
	0 & 0 & -1 & 0\\
	0 & 0 & 0 & -1\\
	1 & 0 & 0 & 0\\
	0 & 1 & 0 & 0\\
\end{bmatrix}
-
\begin{bmatrix}
	b_i & 0 & 0 & 0\\
	0 & b_i & 0 & 0\\
	0 & 0 & 0 & 0\\
	0 & 0 & 0 & 0\\
\end{bmatrix}
\right)
\begin{bmatrix}
	\frac{\partial V_i}{\partial p_{i,x}}\\
	\frac{\partial V_i}{\partial p_{i,y}}\\
	\frac{\partial V_i}{\partial q_{i,x}}\\
	\frac{\partial V_i}{\partial q_{i,y}}
\end{bmatrix}
+
\begin{bmatrix}
	1 & 0 \\
	0 & 1 \\
	0 & 0 \\
	0 & 0 \\
\end{bmatrix}
u_i
\end{equation}
where
\begin{equation}
\label{eq:H_simulations}
V_i(t)=\frac{1}{2m_i}\left(\left(p_{i,x}(t)\right)^2 + \left(p_{i,y}(t)\right)^2\right) + \frac{1}{2}k_i \left(\left(q_{i,x}(t) - \bar{q}_{i,x}\right)^2 + \left(q_{i,y}(t) - \bar{q}_{i,y}\right)^2\right)\,,  
\end{equation}
for $i=1,2,\dots,M$ with $m_i=k_i=1$ and $b_i=0.2$. The inputs $u_i = (F_{i,x}, F_{i,y})$ are forces in the $x$ and $y$ directions. The initial conditions of the systems are fixed and indicated in Figure~\ref{fig:OL_CL} with a ``$\star$'' symbol.

The adjacency matrix $S_c$ of the communication graph $\mathcal{G}_c$ is defined as
\begin{equation}
\label{eq:communication_matrix}
	S_c=
	\begin{bmatrix}
		1 & 1 & 0 & 0 & \dots & 0 & 0 & 1\\
		1 & 1 & 1 & 0 & \dots & 0 & 0 & 0\\
		0 & 1 & 1 & 1 & \dots & 0 & 0 & 0\\
		0 & 0 & 1 & 1 & \dots & 0 & 0 & 0\\
		\vdots & \vdots & \vdots & \vdots & \ddots & \vdots & \vdots \\
		0 & 0 & 0 & 0 & \dots & 1 & 1 & 1 \\
		1 & 0 & 0 & 0 & \dots & 0 & 1 & 1 \\
	\end{bmatrix}
	\in \mathbf{R}^{12\times12}\,.
\end{equation} 
The DeepDisCoPH controller is given by \eqref{eq:controller1}-\eqref{eq:controller2} where
$\blue{\mathbf{R}_c}=\text{blkdiag}(\blue{R_{c,1}},\blue{R_{c,2}},\dots,\blue{R_{c,12}})$, 
$\blue{R_{c,i}} = 12\begin{bsmallmatrix}I_2 & 0 \\0 & 0\end{bsmallmatrix}$
for $i=1\dots,12$. We let $\blue{\mathbf{J}}$ satisfy $\blue{\mathbf{J}}=-\blue{\mathbf{J}}^\top$ and $\blue{\mathbf{K}}$ be trainable matrix parameters that lie in $\text{blkSparse}(S_c)$.

The controller energy is completely separable to comply with $L_\xi = 0$,  and defined as $\Phi(\bm{\xi}(t),\blue{\bm{\theta}(t)}) = \sum_{i=1}^{12}\Phi_i(\xi_i,\blue{\theta_i})$ with $\xi_i\in\mathbb{R}^4$. We define the local energies as
\begin{equation}
	\Phi_i(\xi_i,\blue{\theta}_i) = \tilde{\sigma}(\blue{W_i(t)}\xi_i(t) + \blue{b_i(t)})^\mathsf{T}\mathbf{1}\,,\quad \forall i =1,\ldots,12\,,
\end{equation}
where $\blue{\theta_i} = (\blue{W_i},\blue{b_i})$ and $ \tilde{\sigma}(\cdot)=\log(\cosh(\cdot))$ is applied element-wise.

We implement the training of \eqref{eq:closed_Loop_footnote}-\eqref{eq:closed_Loop} as a Neural ODE endowed with forward Euler discretization. Setting $N=100$ step times, we obtain a time step $h=T/N=0.05$ and $100$ layers for the neural network. 
The trainable parameters $\blue{\mathbf{W}(t)}$ and $\blue{\mathbf{b}(t)}$ are modelled as piece-wise constant functions in each interval of time $[t_k, t_k+h]$ for $t_k = k\cdot h$ and $k=0,1,\dots,N-1$. During training, we fixed $\xi_i(0)=(3,0,0,0)$ for all robots $i$ and simulated trajectories for the same initial conditions $\bm{\zeta}(0)$ for all iterations of gradient descent.

To validate the closed-loop stability properties of DeepDisCoPH controllers in comparison with standard MLPs from \cite{gama2021graph}, we need to simulate the system for $t>T$. For long-horizon simulations, we freeze $\bm{\theta}(t) = \blue{\bm{\theta}(T)}$ for all times $t>T$, and in order to faithfully simulate the continuous-time closed-loop we use the Runge-Kutta~5 integration method. Specifically, all the losses reported in Table~\ref{tab:comparison_MLP} are calculated using Runge-Kutta~5 integration.

We use gradient descent with Adam for $300$ epochs, in order to minimize the loss function
\begin{equation}
    \mathcal{L} + \alpha_{ca}\mathcal{L}_{ca} + \alpha_w\mathcal{R}_w\,,
\end{equation}
with hyperparameters $\alpha_{ca} = 100$ and $\alpha_w = 125-\alpha_{ca}$.
Moreover, for the collision avoidance term, we set a minimum distance $D=0.5$, and for the control loss \eqref{eq:Lx} we set $R(t) = 0.5\,I$ and $Q(t) = \gamma^{T-t} I$ with $\gamma=0.95$ during training and $\gamma=1$ for testing. Here, $\gamma$ acts as a discount factor that assigns less weight to late-horizon costs.


\section{MLP controller}
\label{app:mlp}
We have  compared DeepDisCoPH controllers with an MLP distributed controller as per \cite{gama2021graph}.\footnote{Without the structural assumptions needed for a scalable GNN implementation.}
To keep the comparison fair, we have trained a \textit{time-invariant} DeepDisCoPH — rather than a time-varying one — so that the MLP has a comparable number of training parameters. 

\paragraph{Time-invariant (TI) DeepDisCoPH: implementation details}
In the TI DeepDisCoPHs,  we force parameters to be constant across layers, i.e., $\blue{\mathbf{W}(t)} = \blue{\mathbf{W}}$ and $\blue{\mathbf{b}(t)}=\blue{\mathbf{b}}$ for all $t \in \mathbb{R}$. Moreover, we select the energy function $\Phi_i(\cdot)$ as a 5-layered NN as per
\begin{align*}
    &\omega^0_i = \xi_i(t) \,,\\
    &\omega^{k+1}_i = \operatorname{log}(\operatorname{cosh}(\blue{W^k_i}\omega^k_i + \blue{b^k_i})) \quad \text{for } k=0,\dots,4\,,\\
	&\Phi_i(\xi_i,\blue{\theta_i}) = \omega^{5}_i 1\,.
\end{align*}

\paragraph{MLP implementation details}
The implemented distributed MLP  of Table~\ref{tab:comparison_MLP} is a 3-layered NN with weights $\blue{\mathbf{W^j}} $and bias $\blue{\mathbf{b^j}}$ for $j=0,\dots,2$, and $\tanh(\cdot)$ as activation function.
The layer dimensions at each layer are given by
\begin{equation}
2 \xrightarrow{\blue{W^0_i}} 8 \xrightarrow{\blue{W^1_i}} 8
\xrightarrow{\blue{W^2_i}} 2\,.
\end{equation}
In order to comply with the communication graph given by $S_c$, we set $\mathbf{\blue{W^0}}\in\text{blkSparse}(S_c)$ and $\mathbf{\blue{W^j}},\in\text{blkSparse}(I_{12})$ for $j=1,2$ as suggested in \cite{gama2021graph}.

\vspace{0.3cm}

The results are reported in Table~\ref{tab:comparison_MLP} and Figures \ref{fig:MLP_traj} and \ref{fig:MLP_traj_t}. 
While an MLP controller might achieve a better performance within the training horizon $T$ due to fewer structural assumptions, closed-loop stability cannot be guaranteed even after training. 
Note that this comparison holds both for the previous setting as well as for a much simpler task where
we only minimize $\mathcal{L}_x$, without the collision avoidance task.\footnote{That is, when we set $\alpha_{ca}=0$.}

In confirmation of the closed-loop stability result of Corollary~\ref{pr:CL_stability}, Figure~\ref{fig:OL_CL}(b) shows that using a DeepDisCoPH controller the robots asymptotically reach their target locations for $t>T$. Furthermore,  Table~\ref{tab:comparison_MLP} shows that despite simulating for a longer horizon of $10T$ the cumulative  loss only increases by $3.7\%$. Instead, a trained MLP controller might introduce persistently oscillating trajectories with large amplitude for $t>T$ — as reported in Figure~\ref{fig:MLP_traj} and Figure~\ref{fig:MLP_traj_t}. Correspondingly, the cumulative loss for a horizon of $10T$ may increase by more than $30$ times as we indicate in \textbf{\textcolor{red}{red color}} in Table~\ref{tab:comparison_MLP}.

We last note that, to keep the comparison fair, we have trained the TI DeepDisCoPH controller and the MLP controller using the same optimizer and hyperparameters.  
To improve the performance of the MLP controller, we have trained it for twice as many epochs as the TI DeepDisCoPH controller, i.e., $600$ epochs instead of $300$ epochs.

\begin{minipage}{0.47\linewidth}
\begin{center}
	\includegraphics[width=0.99\linewidth]{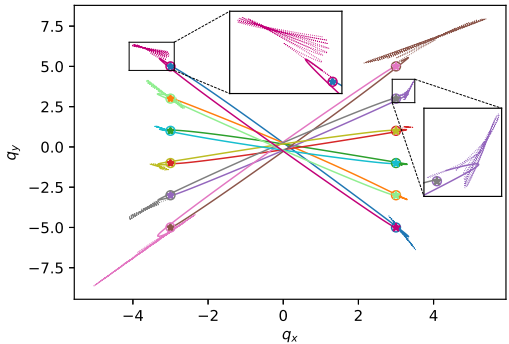}
	\captionof{figure}{Trajectories of the robots in the \textit{xy}-plane when using a distributed MLP controller without c.a. Solid line: training time horizon. Dotted line: extended time horizon.}
	\label{fig:MLP_traj}
\end{center}
\end{minipage}
\hfill
\begin{minipage}{0.47\linewidth}
\begin{center}
	\includegraphics[width=0.99\linewidth]{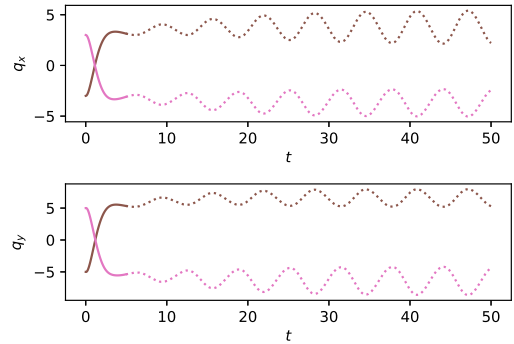}
	\captionof{figure}{Trajectories of the \textit{pink} and \textit{brown} robots when using a distributed MLP controller without c.a. Solid line: training time horizon. Dotted line: extended time horizon.}
	\label{fig:MLP_traj_t}
\end{center}
\end{minipage}

\begin{table}
	\begin{center}
		\caption{Cumulative cost $\mathcal{L}_x$ when comparing a DeepDisCoPH controller with an MLP controller, with and without collision avoidance (c.a.). Training performed for a time horizon of $T=5$.}
		\label{tab:comparison_MLP}
		\small
		\begin{tabular}{c|c|c|c|c}
			& Simulation time & DeepDisCoPH &TI DeepDisCoPH & MLP controller \\
			\hline
			with & $T$ & 34358 & 38355 & 37308 \\ 
			\cline{2-5}
			c.a. & $10\,T$ & 34618 & 39232 & \textbf{\textcolor{red}{1205319}} \\
			\hline
			without & $T$ & 31611 & 31446 & 31433 \\ 
			\cline{2-5}
			c.a. & $10\,T$ & 32769 & 31494 &  \textbf{\textcolor{red}{51133}} \\ 
			\hline
			\hline
			\multicolumn{2}{c|}{\# of trainable parameters} & 24480 & 1680 & 1944 \\
			\hline
		\end{tabular}
	\end{center}
\end{table}


\section{Gradient norms}
\label{app:gradients}

To validate the result of Theorem~\ref{th:vanishing}, we computed the norms of the terms $\frac{\partial \bm{\zeta}_j}{\partial \bm{\zeta}_i}$ for all $i,j$ satisfying $0\leq i < j \leq N$ at every iteration. We verified that gradients never tend to vanish, despite a high number of neural ODE layers and dissipation in the dynamics. Figure~\ref{fig:gradients_N} and Figure~\ref{fig:gradients_N/2} show the sensitivities for the case $j=N$ and $j=N/2$ respectively. As the sensitivity norms do not drop below the level $\approx 10$ despite a large number of neural-network layers, both plots confirm our results of Theorem~\ref{th:vanishing} even if some dissipation is present in the system dynamics. Note also that Figure~\ref{fig:gradients_N} highlights that the most sensitive gradients are with respect to $\zeta_\ell$ for $\ell\in[30,40]$ which coincides with the time interval where the distances between agents decreased. Thus, it is expected that these loss terms are more sensitive to variations. In the case of Figure~\ref{fig:gradients_N/2}, we can appreciate that all gradients are changing during training. This highlights that sensitivity matrices are modified during the learning of optimal trajectories.

\begin{minipage}{0.43\linewidth}
\begin{center}
	\includegraphics[width=0.99\linewidth]{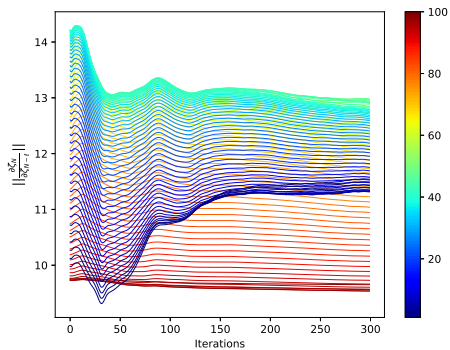}
	\captionof{figure}{Norm of the BSMs ${\frac{\partial \bm{\zeta}_N}{\partial \bm{\zeta}_{N-j}}}$ for $j=1,\dots,100$}
	\label{fig:gradients_N}
\end{center}
\end{minipage}
\hfill
\begin{minipage}{0.43\linewidth}
\begin{center}
    \includegraphics[width=0.99\linewidth]{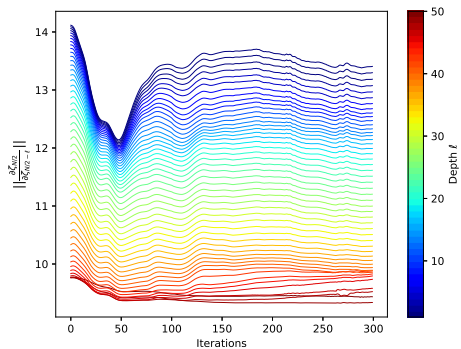}
    \captionof{figure}{Norm of the BSMs ${\frac{\partial \bm{\zeta}_{N/2}}{\partial \bm{\zeta}_{N/2-j}}}$ for $j=1,\dots,50$}
    \label{fig:gradients_N/2}
\end{center}
\end{minipage}


\section{Early stopping of the training}
\label{app:early_stop}

\begin{figure}
\begin{minipage}{0.47\linewidth}
\begin{center}
	\includegraphics[width=0.99\linewidth]{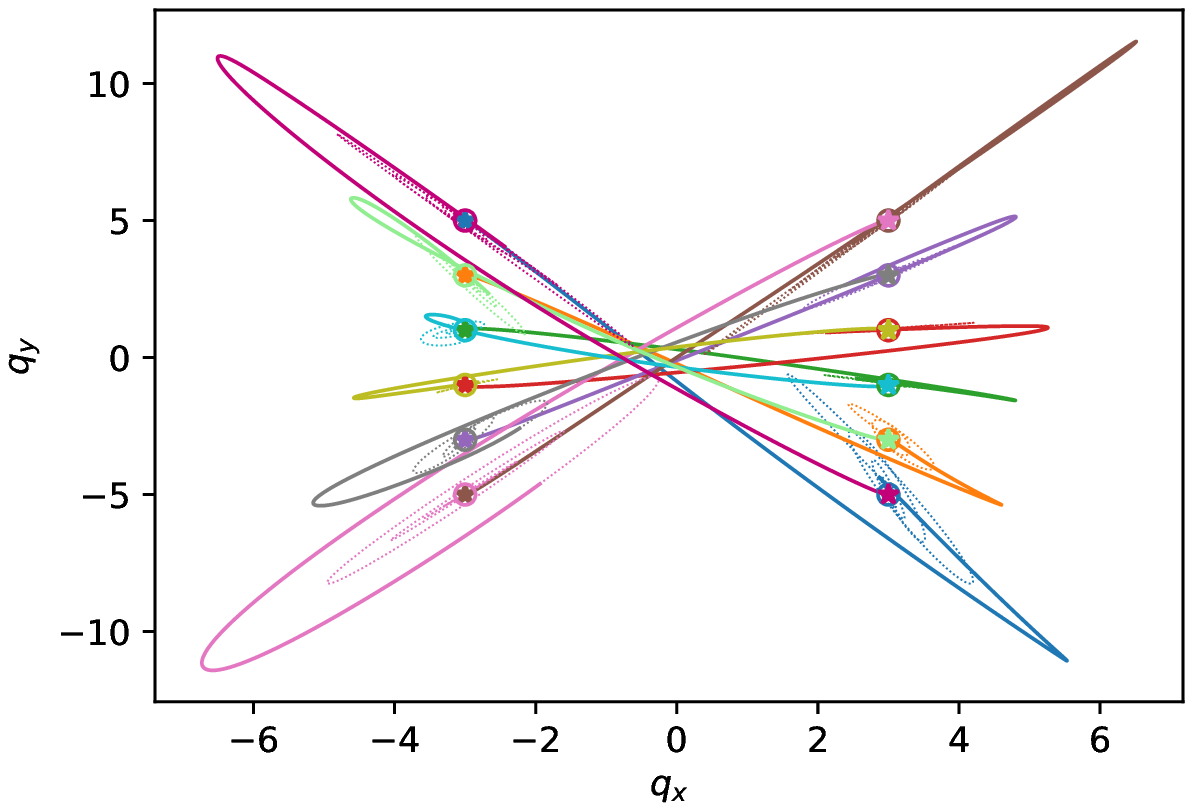}
	(\textit{a})
	\label{fig:train75}
\end{center}
\end{minipage}
\hfill
\begin{minipage}{0.47\linewidth}
\begin{center}
	\includegraphics[width=0.99\linewidth]{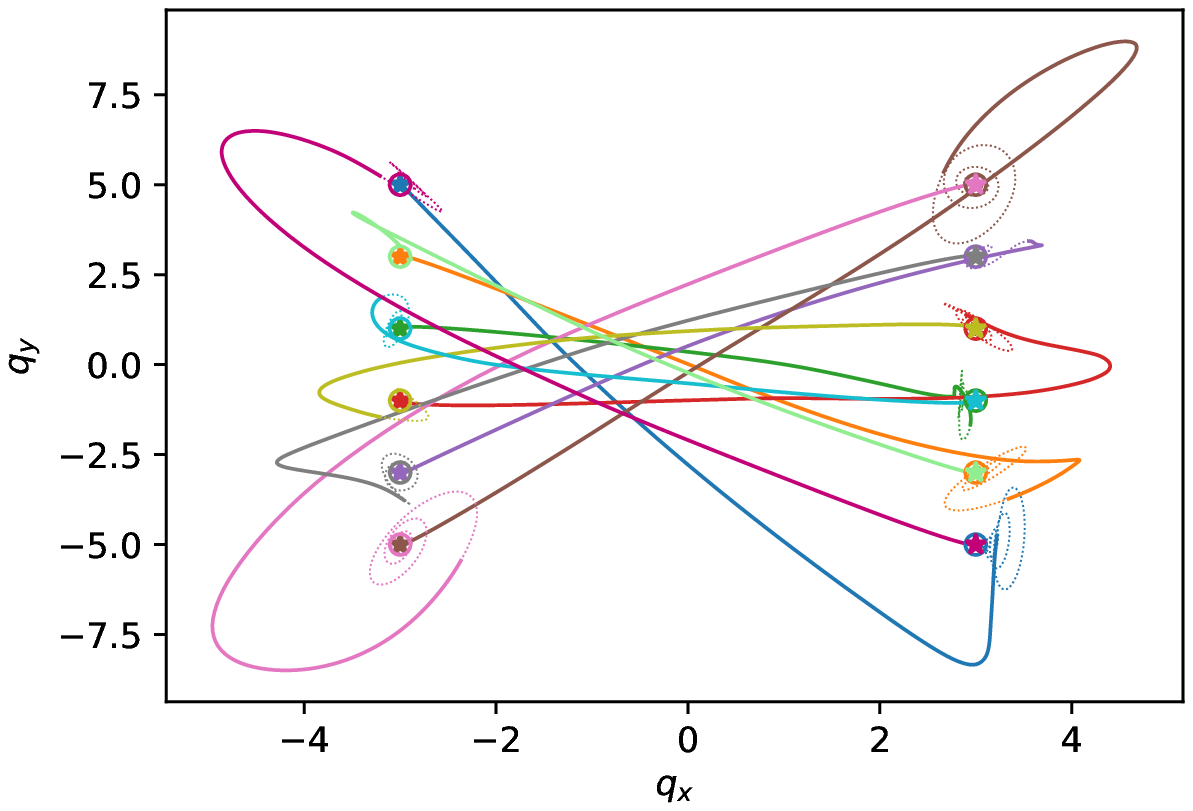}
	(\textit{b})
\end{center}
\end{minipage}

\vspace{16pt}
\begin{minipage}{0.47\linewidth}
\begin{center}
	\includegraphics[width=0.99\linewidth]{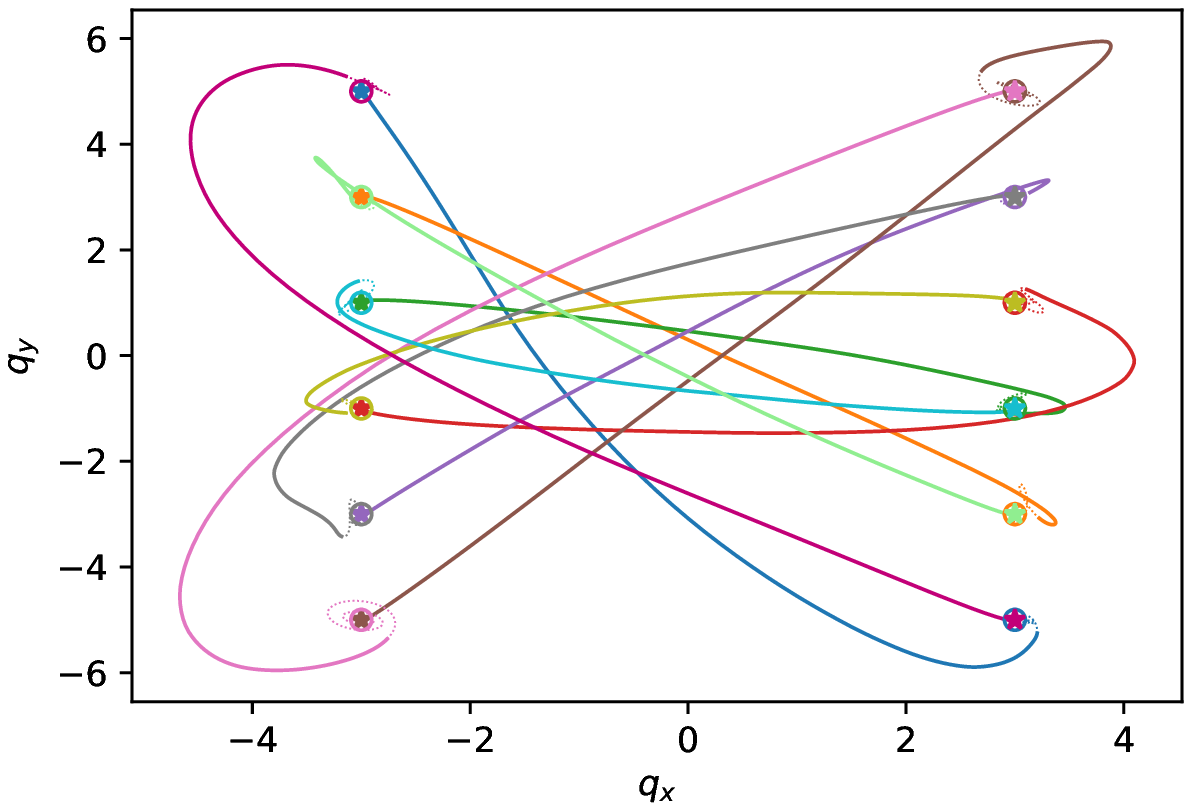}
	(\textit{c})
\end{center}
\end{minipage}
\hfill
\begin{minipage}{0.47\linewidth}
\begin{center}
	\includegraphics[width=0.99\linewidth]{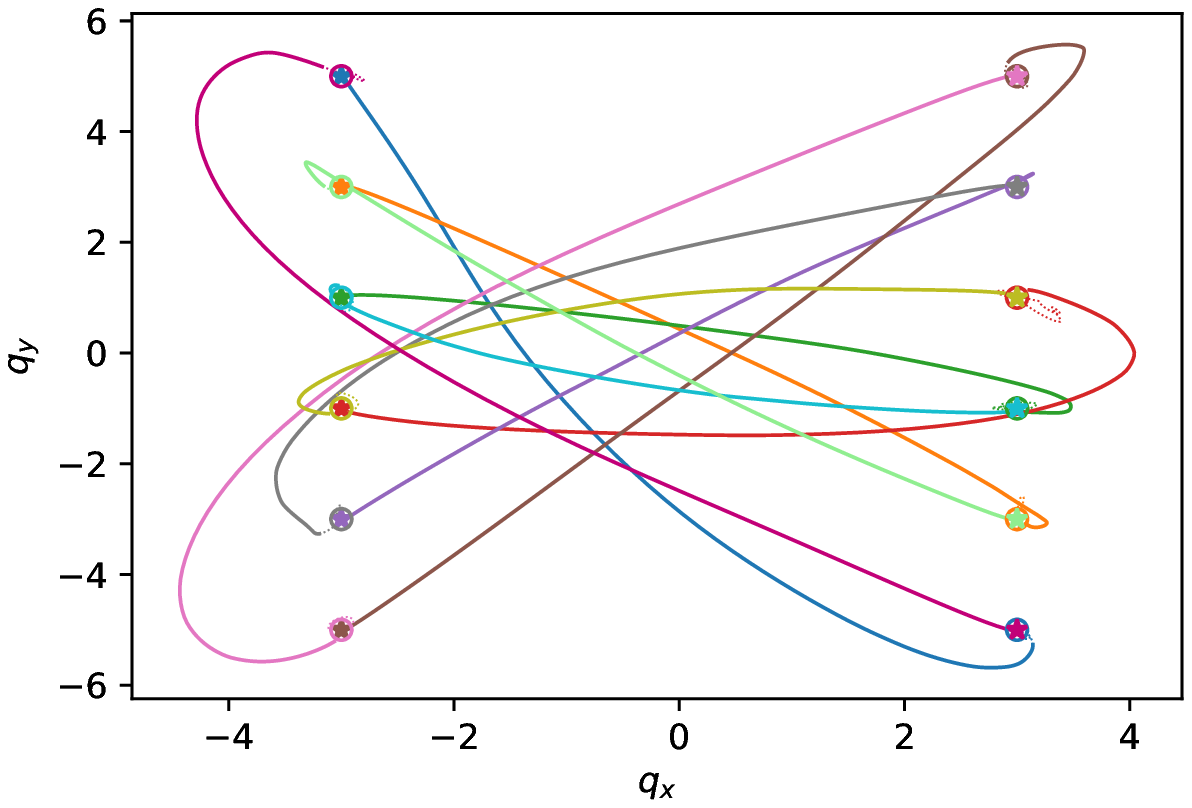}
	(\textit{d})
\end{center}
\end{minipage}
\caption{Robot trajectories in the $xy$-plane when using a partially trained DeepDisCoPH controller with (\textit{a}) 5\%, (\textit{b}) 25\%, (\textit{c}) 50\% and (\textit{d}) 75\% of the total epochs.}
\label{fig:train_stop}
\end{figure}

We verify that DeepDisCoPH controllers ensure closed-loop stability by design even during exploration. 
Specifically, we train the DeepDisCoPH controller for 5\%, 25\%, 50\% and 75\% of the total number of iterations. We report the corresponding closed-loop trajectories of the robots for a long time horizon $t>T$ in Figure~\ref{fig:train_stop}.
In Table~\ref{tab:early_stop}, we report $1)$ the total loss incurred by partially trained controllers before and after the training horizon, and $2)$ the number of collisions. 
As predicted by Corollary~\ref{pr:CL_stability}, partially trained distributed controllers exhibit suboptimal behavior, but never compromise closed-loop system. As expected, the number of collisions decreases drastically during the first iterations of the training, while further training leads to optimized trajectories.

\begin{table}
	\begin{center}
		\caption{Number of collisions and cumulative cost $\mathcal{L}_x$ values when using a partially trained DeepDisCoPH controller, before and after the training horizon of $T=5$ seconds.}
		\label{tab:early_stop}
		\small
		\begin{tabular}{c|c|c|c}
			\% of the training
			& \# of collisions & Loss ($T=5$) & Loss ($T=50$) \\
			\hline
			5\% & 248 & 48809 & 64666 \\ 
			\hline
			25\% & 28 & 38149 & 40078 \\ 
			\hline
			50\% & 2 & 35769 & 36088 \\ 
			\hline
			75\% & 0 & 34980 & 35254 \\ 
			\hline
			100\% & 0 & 34358 & 34618 \\ 
			\hline
		\end{tabular}
	\end{center}
\end{table}

\end{document}